\documentstyle[manuscript,aps]{revtex}
\begin{document}
\preprint{UCONN-96/23;CU-TP-805}

\title{On The Finite Temperature Chern-Simons Coefficient}

\author{Gerald Dunne\footnote{dunne@hep.phys.uconn.edu}}
\address{Physics Department, University of Connecticut, Storrs, CT 06269}
\author{Kimyeong Lee\footnote{klee@phys.columbia.edu} and Changhai
Lu\footnote{chlu@phys.columbia.edu} }
\address{Department of Physics, Columbia University, New York, NY 10027}
\date{\today}
\maketitle

\begin{abstract}
We compute the exact finite temperature effective action in a 0+1-dimensional
field theory containing a topological Chern-Simons term, which has many
features in common with 2+1-dimensional Chern-Simons theories. This exact
result explains the origin and meaning of puzzling temperature dependent
coefficients found in various naive perturbative computations in the higher
dimensional models.
\end{abstract}

\vskip .5in


There are many examples in physics in which the classical Lagrange density
contains a term that is not strictly invariant under a certain transformation
(for example, a `large' gauge transformation), but the classical action changes
by a constant that takes discrete values associated with the `winding number'
of the transformation. For such a system the quantum theory is formally
invariant provided the amplitude $exp(i(action))$ is invariant; thus,
invariance of the quantum theory can be maintained provided the coefficient of
the noninvariant term in the Lagrange density is chosen to take appropriate
discrete values. This argument is familiar in the theory of the Dirac magnetic
monopole, and in Chern-Simons theories\cite{deser}. It is important to ask what
happens to this discretization condition when quantum interactions are taken
into account. For example, the quantum effective action may contain induced
terms of the same noninvariant form, but with a new coefficient. This subject
of induced topological terms is relatively well understood in various examples
of zero temperature quantum field theory\cite{witten,redlich1,pisarski1}.
However, there is currently a great deal of confusion in the corresponding
theories at nonzero temperature.
Typically\cite{niemi,pisarski2,babu,zuk,schaposnik1}, a naive perturbative
computation that mimics the zero temperature computation leads to an induced
topological term equal to the zero temperature induced topological term, but
multiplied by an extra factor of $\tanh(\beta |m|/2)$. Here $\beta=1/T$ is the
inverse temperature, and $m$ is a relevant mass scale. Clearly, this
coefficient cannot take only discrete values for all $T$, even though formal
arguments suggest that it should. This dilemma has recently been
emphasized\cite{pisarski2,schaposnik1} for the particular case of
$2+1$-dimensional fermion and/or Chern-Simons theories, for which quantum
effects may lead to induced Chern-Simons terms. (Related features also appear
in monopole and Aharonov-Bohm systems\cite{parwani,schaposnik2,fosco}). There
is one opinion that anyonic superfluidity should break down at any finite
temperature due to this anomaly\cite{lykken}. There is an opposite opinion that
there is no such temperature dependent anomaly due to some `nonperturbative'
physics. However, we feel that the discussion thus far has missed an essential
point. To illustrate this, we consider a simple analogue of the Chern-Simons
system, which has the advantage that it may be solved exactly and yet it still
retains the essential topological complexities of the problem.

Consider a $0+1$-dimensional field theory of $N_f$ fermions $\psi_j$, $j=1\dots
N_f$, minimally coupled to a $U(1)$ gauge field $A$. It is not possible to
write a Maxwell-like kinetic term for the gauge field in $0+1$-dimensions, but
we can write a Chern-Simons term\footnote{Recall that it is possible to define
a Chern-Simons term in odd dimensional space-time. Some features of
`Chern-Simons quantum mechanics' have been studied previously\cite{dunne}.} -
it is linear in $A$. We formulate the theory in Euclidean space (i.e. imaginary
time) so that we can go smoothly between nonzero and zero temperature using the
imaginary time formalism\cite{kapusta}. The Lagrange density is
\begin{equation}
{\cal L}=\sum_{j=1}^{N_f}\psi^\dagger_j  \left(\partial_\tau-iA+m\right)\psi_j
-i\kappa A
\label{lag}
\end{equation}
There are many similarities between this model and the $2+1$-dimensional model
of fermions coupled to a nonabelian Chern-Simons gauge field. For example, this
model supports gauge transformations with nontrivial winding number. Under the
$U(1)$ gauge transformation $\psi\to e^{i \lambda}\psi$, $ A\to A+\partial_\tau
\lambda$, the Lagrange density changes by a total derivative and the action
changes by
\begin{equation}
\Delta S=-i\kappa\int d \tau\,\partial_\tau \lambda=-2\pi i\kappa N
\label{winding}
\end{equation}
where $N\equiv\frac{1}{2\pi}\int d\tau \partial_\tau \lambda$ is the
integer-valued `winding number' of the topologically nontrivial gauge
transformation. Thus, choosing $\kappa$ to be an integer, the Euclidean action
changes by an integer multiple of $2\pi i$, so that the quantum path integral
is formally invariant -- just as in three dimensional nonabelian Chern-Simons
theories\cite{deser}.

Under naive charge conjugation $C$ : $\psi\to\psi^\dagger$, $A\to -A$, the
fermion mass term and the Chern-Simons term are not invariant. This situation
is similar to the fermion mass term and the Chern-Simons term in three
dimensions, which are not invariant under the parity transformation. In that
case, introducing an equal number of fermions of opposite sign mass, the
fermion mass term can be made invariant under a generalized parity
transformation. Similarly, with an equal number of fermion fields of opposite
sign mass, one can generalize charge conjugation to make the mass term
invariant in our $0+1$-dimensional model.

There is a global part of the $U(1)$ symmetry, whose conserved charge
is
\begin{equation}
Q_F = \frac{1}{2}\sum_j( \psi_j^\dagger \psi_j -\psi_j\psi^\dagger_j)
\label{charge}
\end{equation}
The fermionic Hamiltonian is
\begin{equation}
H_F = \frac{m}{2}\sum_j(\psi_j^\dagger \psi_j - \psi_j\psi^\dagger_j)
 = mQ_F
\label{ham}
\end{equation}
Both $Q_F$ and $H_F$ change sign under charge conjugation. In addition to the
$U(1)$ gauge symmetry, there is a global $SU(N_f)$ flavor symmetry, whose
conserved charges are
\begin{equation}
R^a = \sum_{i,j}\psi^\dagger_i T^a_{ij} \psi_j
\end{equation}
where the $T^a$ are the generators of $SU(N_f)$ in the fundamental
representation.

The canonical commutation relations are $\{\psi_i, \psi_j^\dagger\}
=\delta_{ij}$, and the ground state $|0>$ is chosen so that the energy is
lowest: $\psi_j|0>=0$ if $m>0$, and $\psi_j^\dagger |0>=0$ if
$m<0$. Then the vacuum expectation value of the fermionic charge $Q_F$ at zero
temperature is
\begin{equation}
<0|Q_F|0> = -\frac{m}{2|m|}N_f
\label{vacuum0}
\end{equation}
while at nonzero temperature
\begin{equation}
<0|Q_F|0>_\beta = -\frac{m}{2|m|}N_f \tanh\left(\frac{\beta |m|}{2}\right)
\label{vacuumt}
\end{equation}
Note that the $T=0$ answer (\ref{vacuum0}) is regained smoothly in the zero $T$
($\beta\to\infty$) limit.

We now compute the effective action for this theory:
\begin{equation}
\Gamma[A]=\log\left[{\det\left(\partial_\tau-i A+m\right)\over
\det\left(\partial_\tau+m\right)}\right]^{N_f}
\label{effective}
\end{equation}
Recalling that the fermion fields at finite temperature are {\it antiperiodic},
$\psi(0)=-\psi(\beta)$, the eigenvalues of the operator $\partial_\tau-iA+m$
are
\begin{equation}
\Lambda_n=m-i{a\over \beta}+{(2n-1)\pi i\over \beta}, \qquad\qquad
n=-\infty,\dots,+\infty
\label{eigenvalues}
\end{equation}
where $a\equiv \int_0^\beta d\tau A(\tau)$. To get these, we can make a small
gauge transformation so that $A$ takes the constant value $a/\beta$. Then the
determinants may be computed as usual\cite{jackiw}
\begin{eqnarray}
{\det\left(\partial_\tau-i A+m\right)\over \det\left(\partial_\tau+m\right)}
=\prod_{n=-\infty}^\infty\left[ {m-i{a\over \beta}+{(2n-1)\pi i\over
\beta}\over m+ {(2n-1)\pi i\over \beta}}\right]
= {\cosh\left(\frac{\beta m}{2}-i \frac{a}{2}\right)\over
\cosh\left(\frac{\beta m}{2}\right)}
\label{determinant}
\end{eqnarray}
Thus the exact finite temperature effective action is
\begin{equation}
\Gamma[A]=N_f \log\left[\cos\left(\frac{a}{2}\right)-i \tanh\left(\frac{\beta
m}{2}\right) \sin\left(\frac{a}{2}\right)\right]
\label{answer}
\end{equation}
It is interesting to notice that $\Gamma[A]$ is not an extensive quantity ({\it
i.e.} it is not an integral of a density) in Euclidean time. Rather, it is a
complicated function of the time integral of $A$.

In the zero temperature limit, the tanh function reduces to
$\tanh\left(\frac{\beta m}{2}\right)\to \frac{m}{|m|}$, so that
\begin{equation}
\Gamma[A]_{T=0}=-\frac{i}{2}{m\over |m|}N_f \int_0^\beta d\tau A(\tau)
\label{zero}
\end{equation}
The zero temperature effective action {\it is} an extensive quantity in
Euclidean time. Indeed, $\Gamma[A]_{T=0}$ has the same form as the original
Chern-Simons term, and we conclude that the Chern-Simons coefficient $\kappa$
is shifted in the combined, classical plus effective, action:
\begin{equation}
\kappa\to\kappa-\frac{1}{2}\frac{m}{|m|}N_f
\label{shift}
\end{equation}
The shift $\delta\kappa$ at $T=0$ is exactly the charge (\ref{vacuum0}) of the
$T=0$ fermion ground state, which should dominate the zero temperature
correction. This is quantized in half integer units. (This is the quantum
mechanical analogue of the global anomaly\cite{witten,jackiw}.) If the number
of fermion flavors is even, the vacuum charge is an integer and there is no
global anomaly. This is the same as in $2+1$ dimensional fermion-Chern-Simons
theories.

The finite temperature effective action is more complicated. An expansion of
the exact result (\ref{answer}) in powers of the gauge field yields
\begin{equation}
\Gamma[A]=N_f\left( -\frac{i}{2}\tanh\left(\frac{\beta m}{2}\right)
a-\frac{1}{8} {\rm sech}^2\left(\frac{\beta m}{2}\right) a^2-\frac{i}{24} \tanh
\left(\frac{\beta m}{2} \right) {\rm sech}^2\left(\frac{\beta m}{2}\right) a^3
+\dots\right)
\label{expansion}
\end{equation}
We can compare this exact result with a standard field theoretic perturbative
computation:
\begin{eqnarray}
\Gamma[A]=N_f \log\,\det\left(1-iS A\right)
=-N_f \sum_{p=1}^\infty {i^p\over p}{\rm tr}\left(SASA\dots SA\right)
\label{pert}
\end{eqnarray}
where $S$ is the Green's function for the free operator $(\partial_\tau+m)$.

It is instructive to consider first the case in which $A$ is constant:
$A=a/\beta$. Then
\begin{eqnarray}
{\rm tr}(S^p)&=&\sum_{n=-\infty}^\infty {1\over \left({(2n-1)\pi i\over
\beta}+m\right)^p} =\frac{\beta}{2}{(-1)^{p-1}\over (p-1)!}
\left({\partial\over \partial m} \right)^{p-1} \tanh\left
(\frac{m\beta}{2}\right)
\end{eqnarray}
in which case the perturbative expansion (\ref{pert}) yields the expansion
(\ref{expansion}) of the exact result.

For a general\footnote{Recall that $A(\tau)$ is a periodic function of $\tau$
in the imaginary time formulation of the finite temperature theory.} gauge
field $A(\tau)$ we can compute all orders in the perturbation series
(\ref{pert}) using the Green's function
\begin{eqnarray}
S(\tau-\tau^\prime)&= &\frac{1}{\beta}\sum_{n=-\infty}^\infty {e^{i(2n-1)\pi
i/\beta}\over {(2n-1)\pi i\over \beta}+m}\cr
&=&-\frac{1}{2}\left[ \sinh(m|\tau-\tau^\prime|)-\tanh(\frac{m\beta}{2})
\cosh(m|\tau-\tau^\prime|) \right]\cr
&&\qquad +\frac{1}{2}\epsilon_\beta (\tau-\tau^\prime)
\left[\cosh(m|\tau-\tau^\prime|)- \tanh(\frac{m\beta}{2})
\sinh(m|\tau-\tau^\prime|)\right]
\end{eqnarray}
Here $\epsilon_\beta(\tau)$ is the periodic step function:
\begin{equation}
\epsilon_\beta(\tau)=\cases{+1,\quad 0<\tau<\beta\cr -1,\quad -\beta<\tau<0}
\end{equation}
with $\epsilon_\beta(n\beta)\equiv 0$. Thus $S(0)=\frac{1}{2}
\tanh(\frac{m\beta}{2})$, and $S(\beta)=-\frac{1}{2}\tanh(\frac{m\beta}{2})$;
while for $0<\tau<\beta$, $S(\tau)=\frac{1}{2}(1+\tanh(\frac{m\beta}{2}))
e^{-m\tau}$, and for $-\beta<\tau<0$, $S(\tau)=-\frac{1}{2}(1-\tanh(
\frac{m\beta}{2})) e^{-m\tau}$. Equipped with these results for the Green's
function we can now compute the $p^{th}$ order contribution to the perturbative
expansion (\ref{pert}) of the effective action:
\begin{eqnarray}
&&-{i^p\over p}\int_0^\beta d\tau_1 \int_0^\beta d\tau_2\dots \int_0^\beta
d\tau_p S(\tau_1-\tau_2)A(\tau_2)S(\tau_2-\tau_3)A(\tau_3)\dots
S(\tau_p-\tau_1)A(\tau_1)\cr
&&\hskip .5in ={(-i)^p\over 2^p p!} \, \left[\left({\partial\over
\partial(\frac{m\beta}{2})}\right)^{p-1} \tanh(\frac{m\beta}{2})\right]\,
\left[\int_0^\beta A(\tau)d\tau\right]^p
\end{eqnarray}
We obtain the same expansion as the expansion (\ref{expansion}) of the exact
answer (\ref{answer}). Once again we see that the full effective action
$\Gamma[A]$ is not an extensive quantity in Euclidean time.

We can alternatively derive the finite temperature effective action from the
partition function in a grand canonical ensemble. Consider a constant gauge
field $A(\tau)=-i\mu$. Then
\begin{equation}
e^{\Gamma[A]} = \frac{{\rm tr}\,\exp [-\beta H_F +\beta\mu Q_F]}{{\rm tr}\,\exp
[-\beta H_F ]}
\end{equation}
where $H_F$ is the fermionic Hamiltonian (\ref{ham}) and $Q_F$ is the fermionic
charge (\ref{charge}). A straightforward calculation yields
\begin{equation}
e^{\Gamma[A]} = \left[ \frac{\cosh\frac{\beta(m-\mu)}{2}} {\cosh\frac{\beta
m}{2}} \right]^{N_f}
\end{equation}
in agreement with (\ref{determinant},\ref{answer}).

Our $0+1$-dimensional model is special in the sense that we are able to compute
{\it every order} in perturbation theory, and furthermore we are able to {\it
re-sum} the perturbative expansion to obtain the {\it exact} effective action.
In higher dimensional examples it is generally not possible to compute the
exact effective action, because of the extra momentum integrations and the
additional tensor or spinor structure. Thus, a field theoretic computation of
the finite temperature effective action in
$2+1$-dimensions\cite{niemi,pisarski2,babu,zuk,schaposnik1} is a perturbative
one that typically looks only for the {\it lowest order} term, which happens to
be the same as the original Chern-Simons term. Applying this same philosophy to
the example treated here, we would (erroneously) conclude that the effective
action is just the first term in the expansion (\ref{expansion}), and hence
that the Chern-Simons coefficient is shifted by a temperature-dependent amount
\begin{equation}
\kappa\to\kappa-\frac{1}{2}\frac{m}{|m|}\tanh(\frac{\beta |m|}{2}) N_f
\label{tshift}
\end{equation}
This temperature dependent shift $\delta\kappa$ is precisely the result
obtained\cite{niemi,pisarski2,babu,zuk,schaposnik1} in fermion and/or
Chern-Simons theory in $2+1$-dimensions. There is an obvious physical
interpretation of the correction in (\ref{tshift}) : this shift is just the
finite temperature expectation value (\ref{vacuumt}) of the fermion charge. In
the three dimensional theory the corresponding correction to $\kappa$ can be
interpreted as the induced charge density per magnetic field. In the high
temperature limit, the expectation value of the fermion charge goes to zero as
the energy gap between the excited states and the ground state is negligible in
this limit.

Attempting to identify the shifted coefficient in (\ref{tshift}) as a new
Chern-Simons coefficient, which should take discrete values in its own right,
leads immediately to the difficulties discussed in the introduction and in
Refs.\cite{pisarski2,schaposnik1}. However, as is very clear from our exactly
solvable model, such an identification is {\it incorrect} at finite temperature
because it ignores the higher terms in the perturbative expansion of the
effective action. At zero temperature it happens to be correct to make this
identification because only the first term in the expansion (\ref{expansion})
survives in the zero temperature limit. Indeed, we see from the exact finite
$T$ effective action (\ref{answer}) that the entire effective action has a
well-defined behavior under a large gauge transformation, {\it independent of
the temperature}, even though at any given finite order of a perturbation
expansion there is a temperature dependence. Under a large gauge
transformation, $a\to a+2\pi N$, the effective action (\ref{answer}) is shifted
by $N\pi i$, which is exactly the same behavior as at zero temperature -- see
(\ref{zero}) and (\ref{shift}). This global flavor anomaly may be avoided,
independent of the temperature, by considering an even number of fermion
flavors. However, if the effective action is computed to any finite order in
perturbation theory, its transformation under a large gauge transformation is
complicated and temperature dependent.

This simple model implies that discussion of the gauge invariance of finite
temperature effective actions and induced Chern-Simons terms in higher
dimensions requires, at the very least, consideration of the full perturbation
series. Conversely, no sensible conclusion may be drawn by considering only the
first term in the expansion, as previous work has attempted to do. Our work
suggests that once we remove the global flavor anomaly we expect the entire
finite temperature effective action to be invariant under large gauge
transformations.

An interesting feature of this model is that the finite temperature effective
action is not an extensive quantity in Euclidean time. While we expect an
effective action to be an extensive quantity in space, there is no reason why
it should be so in Euclidean time. We expect that in the three dimensional
calculation of the finite temperature effective action we could expand in the
spatial derivatives and spatial components of the gauge field, but we would
need to keep {\it all} terms in time integrations and time derivatives and in
$A_0$. This requirement explains why the standard argument for gauge invariance
of just the Chern-Simons-like term in the effective action, based on an
arbitrary scaling of large gauge transformations, works at zero $T$ but fails
at nonzero $T$\cite{pisarski2}. Nevertheless, the leading order term in a
spatial expansion should itself be invariant under large gauge transformations.
It would be interesting to find the exact expression for this effective action
and its physical meaning.

Finally, the Chern-Simons quantum mechanics model considered here may be
generalized to incorporate also bosonic degrees of freedom, with conserved
$U(1)$ charge $Q_B$. Then the Gauss law constraint becomes $\kappa+Q_F+Q_B=0$,
which leads to interesting superselection sectors of integer total charge. In
addition, one can introduce Yukawa couplings between the bosonic and fermionic
fields, and supersymmetrize the system. We believe further investigation in
this direction will be rewarding.

\vskip .5in
\noindent{\bf Acknowledgements}

Each author thanks the US Department of Energy for support. KL is supported by
an NSF Presidential Young Investigator Fellowship.

\end{document}